# FCC structured ferromagnetic ultra-thin films with two spin layers described by fourth order perturbed Heisenberg Hamiltonian

*M.S.M. Farhan*

Department of Physical Sciences, South Eastern University of Sri Lanka, Sammanthurai, Sri Lanka.

**Abstract**

Fourth order perturbation was applied to study a small variation of the azimuthal angle of spin of fcc structured ferromagnetic thin films with two spin layers. The varaiation of magnetic energy and the orientation of magnetic easy axis with the fourth order magnetic anisotropy conatant in two spin layers was investigated. When the second order magnetic anisotropy constant at the top spin layer is higher than that of bottom spin layer, the total magnetic energy is slightly higher. Some spikes appeared in the 2D plot of of magnetic energy versus azimuthal angle of spin. According to the 3D plots, the peak value of the magnetic energy gradually decreases with the increase of the stress induced anisotropy constant, and thereafter the peak value of the magnetic energy gradually increases with the increase of the stress induced anisotropy constant. The angle between magnetic easy and hard directions was not 90 degrees. The magnetic easy and hard directions of the film with a higher second order magnetic anisotropy constant of top layer are different from the magnetic easy and hard directions of the film with a lower second order magnetic anisotropy constant of top layer. When the second order magnetic anisotropy constant of the bottom layer is increased, the total magnetic enerhy does not change.

**1. Introduction:**

Ferromagnetic films are prime candidates of magnetic memory and microwave devices. The orientation of easy and hard direction is important in all these applications. Ferromagnetic films have been described using many different models. The quasistatic magnetic hysteresis of ferromagnetic thin films grown a vicinal substrate has been theoretically investigated using Monte carlo simulations [1]. Magnetic properties of ferromagnetic thin films with alternating super layer were studied by Ising model [2]. Structural and magnetic properties of two dimensional FeCo ordered alloys have been determined by first principles band structure theory [3]. Magnetic layers of Ni on Cu have been theoretically studied by using the Korringa-Kohn-Rostoker Green's fuction method [4].

The experimental values of magnetic moments of ferromagnetic materials are different from the theoretical due to the overlapping sub-shells. Magnetic thin films are heated and subsequently cooled during the annealing process. During the heating and cooling process, the stress induced magnetic anisotropy ($K_s$) occurs due to the difference between the thermal expansion coefficients in magnetic film and substrate. For soft magnetic materials, the stress induced anisotropy is comparable to the magneto crystalline anisotropy. Therefore, the coercivity depends on the stress induced anisotropy [5]. The magnetic properties of ferromagnetic and ferrite thin and thick films have been investigated by us [6-14].

In this manuscript, the total energy of ferromagnetic thin films will be described by solving the classical Heisenberg Hamiltonian. In this approach, all energy terms such as magnetic energy, spin dipole interaction energy, spin exchange interaction energy, second and fourth order anisotropy, stress induced anisotropy and demagnetization factor terms will be considered. Fourth order perturbed Heisenberg Hamiltonian with all seven magnetic energy parameters are explained for fcc structure. MATLAB computer software was used to plot 3D and 2D graph of energy versus stress induced anisotropy and azimuthal angle of spin.

## 2. Model:

The Heisenberg Hamiltonian of ferromagnetic films can be formulated as following [8-10].

$$H = -\frac{J}{2}\sum_{m,n}\vec{S}_m \cdot \vec{S}_n$$

$$+\frac{\omega}{2}\sum_{m\neq n}\left(\frac{\vec{S}_m \cdot \vec{S}_n}{r_{mn}^3} - \frac{3(\vec{S}_m \cdot \vec{r}_{mn})(\vec{r}_{mn} \cdot \vec{S}_n)}{r_{mn}^5}\right) - \sum_m D_{\lambda_m}^{(2)}(S_m^z)^2$$

$$-\sum_m D_{\lambda_m}^{(4)}(S_m^z)^4$$

$$-\sum_{m,n}[\vec{H} - (N_d\vec{S}_n/\mu_0)]\cdot\vec{S}_m - \sum_m K_s \sin 2\theta_m$$

Here $\vec{S}_m$ and $\vec{S}_n$ are two spins. Above equation can be simplified into following form

$$E(\theta) = -\frac{1}{2}\sum_{m,n=1}^{N}\left[\left(JZ_{|m-n|} - \frac{\omega}{4}\Phi_{|m-n|}\right)\cos(\theta_m - \theta_n) - \frac{3\omega}{4}\Phi_{|m-n|}\cos(\theta_m + \theta_n)\right]$$

$$-\sum_{m=1}^{N}(D_m^{(2)}\cos^2\theta_m + D_m^{(4)}\cos^4\theta_m + H_{in}\sin\theta_m + H_{out}\cos\theta_m)$$

$$+ \sum_{m,n=1}^{N} \frac{N_d}{\mu_0} \cos(\theta_m - \theta_n) - K_s \sum_{m=1}^{N} \sin 2\theta_m \qquad (1)$$

Here $N$, $m$ (or $n$), $J$, $Z_{|m-n|}$, $\omega$, $\Phi_{|m-n|}$, $\theta_m(\theta_n)$, $D_m^{(2)}$, $D_m^{(4)}$, $H_{in}$, $H_{out}$, $N_d$ and $K_s$ are total number of layers, layer index, spin exchange interaction, number of nearest spin neighbors, strength of long range dipole interaction, partial summations of dipole interaction, azimuthal angles of spins, second and fourth order anisotropy constants, in plane and out of plane applied magnetic fields, demagnetization factor and stress induced anisotropy constants respectively.

The spin structure is considered to be slightly disoriented. Therefore, the spins could be considered to have angles distributed about an average angle $\theta$. By choosing azimuthal angles as

$\theta_m = \theta + \varepsilon_m$ and $\theta_n = \theta + \varepsilon_n$

Where the $\varepsilon$'s are small positive or negative angular deviations.

Then, $\theta_m - \theta_n = \varepsilon_m - \varepsilon_n$ and $\theta_m + \theta_n = 2\theta + \varepsilon_m + \varepsilon_n$. After substituting these new angles in above equation number (1), the cosine and sine terms can be expanded up to the fourth order of $\varepsilon_m$ and $\varepsilon_n$ as following.

$$E(\theta) = E_0 + E(\varepsilon) + E(\varepsilon^2) + E(\varepsilon^3) + E(\varepsilon^4) + \ldots\ldots\ldots..$$

If the fifth and higher order perturbations are neglected, then

$$E(\theta) = E_0 + E(\varepsilon) + E(\varepsilon^2) + E(\varepsilon^3) + E(\varepsilon^4) \qquad (2)$$

Here

$$E_0 = -\frac{1}{2} \sum_{m,n=1}^{N} \left( J Z_{|m-n|} - \frac{\omega}{4} \Phi_{|m-n|} \right) + \frac{3\omega}{8} \cos 2\theta \sum_{m,n=1}^{N} \Phi_{|m-n|} - \cos^2\theta \sum_{m=1}^{N} D_m^{(2)}$$

$$-\cos^4\theta \sum_{m=1}^{N} D_m^{(4)} - N(H_{in}\sin\theta + H_{out}\cos\theta + K_s\sin 2\theta)$$

$$+ \frac{N_d N^2}{\mu_0} \qquad (3)$$

$$E(\varepsilon) = -\frac{3\omega}{8} sin2\theta \sum_{m,n=1}^{N} \Phi_{|m-n|} (\varepsilon_m + \varepsilon_n) + sin2\theta \sum_{m=1}^{N} D_m^{(2)} \varepsilon_m$$

$$+ 2cos^2\theta sin2\theta \sum_{m=1}^{N} D_m^{(4)} \varepsilon_m$$

$$- H_{in} cos\theta \sum_{m=1}^{N} \varepsilon_m + H_{out} sin\theta \sum_{m=1}^{N} \varepsilon_m$$

$$- 2K_s cos2\theta \sum_{m=1}^{N} \varepsilon_m \qquad (4)$$

$$E(\varepsilon^2) = \frac{1}{4} \sum_{m,n=1}^{N} \left(JZ_{|m-n|} - \frac{\omega}{4} \Phi_{|m-n|}\right) (\varepsilon_m - \varepsilon_n)^2 - \frac{3\omega}{16} cos2\theta \sum_{m,n=1}^{N} \Phi_{|m-n|} (\varepsilon_m + \varepsilon_n)^2$$

$$+ cos2\theta \sum_{m=1}^{N} D_m^{(2)} \varepsilon_m^2 + 2cos^2\theta(cos^2\theta - 3sin^2\theta) \sum_{m=1}^{N} D_m^{(4)} \varepsilon_m^2 + \frac{H_{in}}{2} sin\theta \sum_{m=1}^{N} \varepsilon_m^2$$

$$+ \frac{H_{out}}{2} cos\theta \sum_{m=1}^{N} \varepsilon_m^2 - \frac{N_d}{2\mu_0} \sum_{m,n=1}^{N} (\varepsilon_m - \varepsilon_n)^2$$

$$+ 2K_s sin2\theta \sum_{m=1}^{N} \varepsilon_m^2 \qquad (5)$$

$$E(\varepsilon^3) = \frac{\omega}{16} sin2\theta \sum_{m,n=1}^{N} \Phi_{|m-n|} (\varepsilon_m + \varepsilon_n)^3 - \frac{4}{3} sin\theta cos\theta \sum_{m=1}^{N} D_m^{(2)} \varepsilon_m^3$$

$$- 4sin\theta cos\theta \left(\frac{5}{3} cos^2\theta - sin^2\theta\right) \sum_{m=1}^{N} D_m^{(4)} \varepsilon_m^3 + \frac{H_{in}}{6} cos\theta \sum_{m=1}^{N} \varepsilon_m^3$$

$$- \frac{H_{out}}{6} sin\theta \sum_{m=1}^{N} \varepsilon_m^3$$

$$+ \frac{4}{3} K_s cos2\theta \sum_{m=1}^{N} \varepsilon_m^3 \qquad (6)$$

$$E(\varepsilon^4) = -\frac{1}{48}\sum_{m,n=1}^{N}\left(JZ_{|m-n|} - \frac{\omega}{4}\Phi_{|m-n|}\right)(\varepsilon_m - \varepsilon_n)^4$$

$$+ \frac{\omega}{64}\cos2\theta \sum_{m,n=1}^{N} \Phi_{|m-n|}(\varepsilon_m + \varepsilon_n)^4$$

$$-\frac{1}{3}\cos2\theta \sum_{m=1}^{N} D_m^{(2)}\varepsilon_m^4 - \left(\frac{5}{3}\cos^4\theta - 8\cos^2\theta\sin^2\theta + \sin^4\theta\right)\sum_{m=1}^{N} D_m^{(4)}\varepsilon_m^4$$

$$-\frac{H_{in}}{24}\sin\theta \sum_{m=1}^{N}\varepsilon_m^4 - \frac{H_{out}}{24}\cos\theta \sum_{m=1}^{N}\varepsilon_m^4 + \frac{N_d}{24\mu_0}\sum_{m,n=1}^{N}(\varepsilon_m - \varepsilon_n)^4$$

$$-\frac{2}{3}K_s\sin2\theta \sum_{m=1}^{N}\varepsilon_m^4 \tag{7}$$

For films with two spin layers, $N = 2$. Therefore, $m$ and $n$ change from 1 to 2.

$$E_0 = -JZ_0 + \frac{\omega}{4}\Phi_0 - JZ_1 + \frac{\omega}{4}\Phi_1 + \frac{3\omega}{4}\cos2\theta(\Phi_0 + \Phi_1) - \cos^2\theta\left(D_1^{(2)} + D_2^{(2)}\right)$$

$$-\cos^4\theta\left(D_1^{(4)} + D_2^{(4)}\right) - 2(H_{in}\sin\theta + H_{out}\cos\theta + K_s\sin2\theta)$$

$$+ \frac{4N_d}{\mu_0} \tag{8}$$

$$E(\varepsilon) = -\frac{3\omega}{4}\sin2\theta[(\Phi_0 + \Phi_1)(\varepsilon_1 + \varepsilon_2)] + \sin2\theta\left(D_1^{(2)}\varepsilon_1 + D_2^{(2)}\varepsilon_2\right)$$

$$+ 2\cos^2\theta\sin2\theta\left(D_1^{(4)}\varepsilon_1 + D_2^{(4)}\varepsilon_2\right) - H_{in}\cos\theta(\varepsilon_1 + \varepsilon_2) + H_{out}\sin\theta(\varepsilon_1 + \varepsilon_2)$$

$$-2K_s\cos2\theta(\varepsilon_1 + \varepsilon_2) \tag{9}$$

$$E(\varepsilon^2) = \frac{1}{2}\left(JZ_1 - \frac{\omega}{4}\Phi_1\right)(\varepsilon_1 - \varepsilon_2)^2 - \frac{3\omega}{8}\cos2\theta[2\Phi_0(\varepsilon_1^2 + \varepsilon_2^2) + \Phi_1(\varepsilon_1 + \varepsilon_2)^2]$$

$$+ \cos2\theta\left(D_1^{(2)}\varepsilon_1^2 + D_2^{(2)}\varepsilon_2^2\right) + 2\cos^2\theta(\cos^2\theta - 3\sin^2\theta)\left(D_1^{(4)}\varepsilon_1^2 + D_2^{(4)}\varepsilon_2^2\right)$$

$$+ \frac{H_{in}}{2}\sin\theta(\varepsilon_1^2 + \varepsilon_2^2) + \frac{H_{out}}{2}\cos\theta(\varepsilon_1^2 + \varepsilon_2^2) - \frac{N_d}{\mu_0}(\varepsilon_1 - \varepsilon_2)^2$$

$$+ 2K_s\sin2\theta(\varepsilon_1^2 + \varepsilon_2^2) \tag{10}$$

$$E(\varepsilon^3) = \frac{\omega}{8}\sin2\theta[4\Phi_0(\varepsilon_1^3 + \varepsilon_2^3) + \Phi_1(\varepsilon_1 + \varepsilon_2)^3] - \frac{4}{3}\sin\theta\cos\theta\left(D_1^{(2)}\varepsilon_1^3 + D_2^{(2)}\varepsilon_2^3\right)$$

$$-4\sin\theta\cos\theta\left(\frac{5}{3}\cos^2\theta - \sin^2\theta\right)\left(D_1^{(4)}\varepsilon_1^3 + D_2^{(4)}\varepsilon_2^3\right) + \frac{H_{in}}{6}\cos\theta(\varepsilon_1^3 + \varepsilon_2^3)$$

$$-\frac{H_{out}}{6}\sin\theta(\varepsilon_1^3 + \varepsilon_2^3)$$

$$+\frac{4}{3}K_s\cos2\theta(\varepsilon_1^3 + \varepsilon_2^3) \qquad (11)$$

$$E(\varepsilon^4) = -\frac{1}{24}\left(JZ_1 - \frac{\omega}{4}\Phi_1\right)(\varepsilon_1 - \varepsilon_2)^4 + \frac{\omega}{32}\cos2\theta[8\Phi_0(\varepsilon_1^4 + \varepsilon_2^4) + \Phi_1(\varepsilon_1 + \varepsilon_2)^4]$$

$$-\frac{1}{3}\cos2\theta\left(D_1^{(2)}\varepsilon_1^4 + D_2^{(2)}\varepsilon_2^4\right) - \left(\frac{5}{3}\cos^4\theta - 8\cos^2\theta\sin^2\theta\right.$$

$$\left.+\sin^4\theta\right)\left(D_1^{(4)}\varepsilon_1^4 + D_2^{(4)}\varepsilon_2^4\right) - \frac{H_{in}}{24}\sin\theta(\varepsilon_1^4 + \varepsilon_2^4) - \frac{H_{out}}{24}\cos\theta(\varepsilon_1^4 + \varepsilon_2^4)$$

$$+\frac{N_d}{12\mu_0}(\varepsilon_1 - \varepsilon_2)^4 - \frac{2}{3}K_s\sin2\theta(\varepsilon_1^4 + \varepsilon_2^4) \qquad (12)$$

First ($\alpha$), second ($C$), third ($\beta$) and fourth ($F\ and\ G$) order perturbation term can be found in terms of a row and column matrix with all seven terms. Then, the total magnetic energy in equation (2) can be deduced to

$$E(\theta) = E_0 + \vec{\alpha}.\vec{\varepsilon} + \frac{1}{2}\vec{\varepsilon}.C.\vec{\varepsilon} + \varepsilon^2.\beta.\vec{\varepsilon} + \varepsilon^3.F.\vec{\varepsilon} + \varepsilon^2.G.\varepsilon^2 \qquad (13)$$

For the minimum energy of the second order perturbed term

$$\vec{\varepsilon} = -C^+.\alpha \qquad (14)$$

Here $C^+$ is the pseudo inverse of matrix C. $C^+$ can be found using

$$C.C^+ = 1 - \frac{E}{N}$$

Here $E$ is the matrix with all elements given by $E_{mn} = 1$. $I$ is the identity matrix.

Therefore,

$$C_{11}^+ = C_{22}^+$$
$$= \frac{C_{21} + C_{22}}{2(C_{11}C_{22} - C_{21}^2)} \qquad (15)$$

$$C_{12}^+ = C_{21}^+$$

$$= \frac{C_{21} + C_{22}}{2(C_{21}^2 - C_{11}C_{22})} \tag{16}$$

Therefore, from the matrix equation (27)

$$\varepsilon_1 = (\alpha_2 - \alpha_1)\, C_{11}^+ \tag{17}$$

$$\varepsilon_2 = (\alpha_2 - \alpha_1)\, C_{21}^+ \tag{18}$$

After substituting ε in equation (13), the total magnetic energy can be determined.

## 3. Results and Discussion:

All the graphs in this manuscript were plotted for ferromagnetic films with face center cubic lattice and two spin layers. For ferromagnetic films with fcc(001) structure, $Z_0=4$, $Z_1=4$, $Z_2=0$, $\Phi_0 = 9.0336$ and $\Phi_1 = 1.4294$ [15-17]. 3D plot of energy versus angle and stress induced anisotropy constant is given in figure 1 for $\frac{D_1^{(4)}}{\omega} = 5$ and $\frac{D_2^{(4)}}{\omega} = 10$ and figure 3 for $\frac{D_1^{(4)}}{\omega} = 10$ and $\frac{D_2^{(4)}}{\omega} = 5$. Here other parameters are fixed at $\frac{J}{\omega} = \frac{H_{in}}{\omega} = \frac{H_{out}}{\omega} = \frac{N_d}{\mu_0 \omega} = \frac{D_2^{(2)}}{\omega} = \frac{D_1^{(2)}}{\omega} = 10$ for this simulation. The energy in this graph is in the order of $10^{51}$. The same order of energy was observed in bcc structure using fourth order perturbed Heisenberg Hamiltonian with all seven magnetic parameters [18]. The energy maximum can be observed at $\frac{K_s}{\omega} = 12, 18, 28, 35$ and $90$. The major maximum was observed at about $\frac{K_s}{\omega} = 12$. Minimum value of energy is zero. Compared to the 3D plots obtained using third order perturbed Heisenberg Hamiltonian, peaks are closely packed in 3D plots obtained using fourth order perturbed Heisenberg Hamiltonian [11].

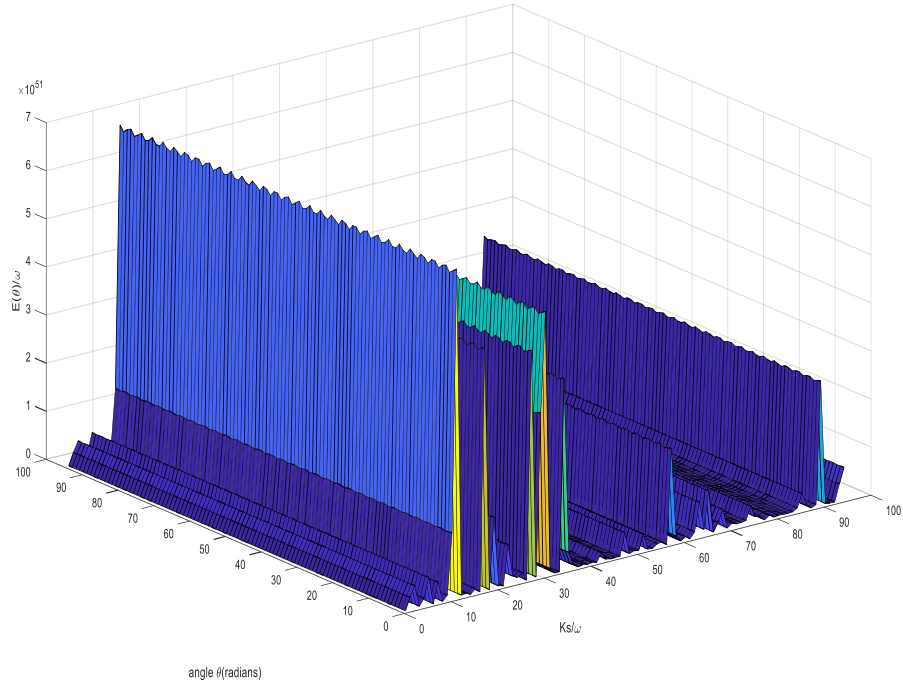

**Figure 1: 3-D plot of energy versus angle and stress induced anisotropy constant for $\frac{D_1^{(4)}}{\omega} = 5$, $\frac{D_2^{(4)}}{\omega} = 10$**

Figure 2 shows the graph of energy versus angle for $\frac{K_S}{\omega} = 12$. Here other parameters were kept at the values given above. The energy minimums can be observed at 0.6283 and 3.581 radians. The major minimum was observed at about 0.6283 radians. The energy maximums can be found at 2.419 and 5.089 radians. The major maximum was observed at about 5.089 radians. Therefore, the angle between consecutive magnetic easy and hard direction is not 90 degrees.

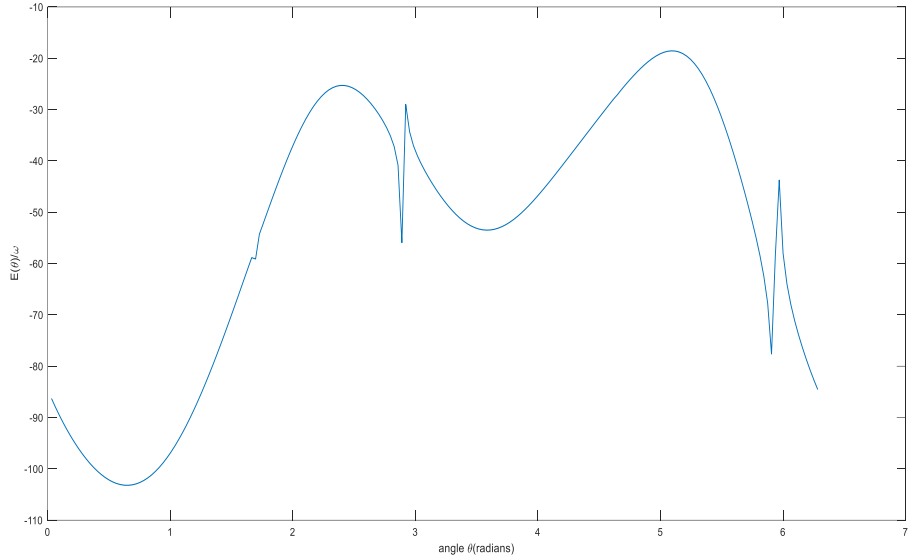

**Figure 2: Graph of energy versus angle for $\frac{K_S}{\omega} = 12$**

As shown in figure 3, several peaks can be observed. However, the gap between two peaks is not a constant. In this graph, the energy maximum can be observed at $\frac{K_S}{\omega} = 5, 15, 48, 52$ and $76$. The major maximum was observed at about $\frac{K_S}{\omega} = 15$. Minimum value of energy is zero. The energy is in the order of $10^{49}$. According to figure 1 and 3, when the fourth order magnetic anisotropy of bottom layer is higher than that of the top spin layer, the order of magnetic energy slightly decreases.

Figure 4 shows the graph of energy versus angle for $\frac{K_S}{\omega} = 15$. The energy minimums can be observed at 0.6283 and 3.644 radians. The major minimum was observed at about 0.6283 radians. The energy maximums can be found at 2.513 and 5.184 radians. The major maximum was observed at about 5.184 radians. The angle between consecutive magnetic easy and hard direction is not 90 degrees in this case too. For the ferromagnetic films with two spin layers in sc structure, the angle between consecutive magnetic easy and hard direction was exactly 90 degrees [19]. The spike observed in this plot did not appear in the graphs obtained for bcc structure with same spin layers using fourth order perturbed Heisenberg Hamiltonian [18].

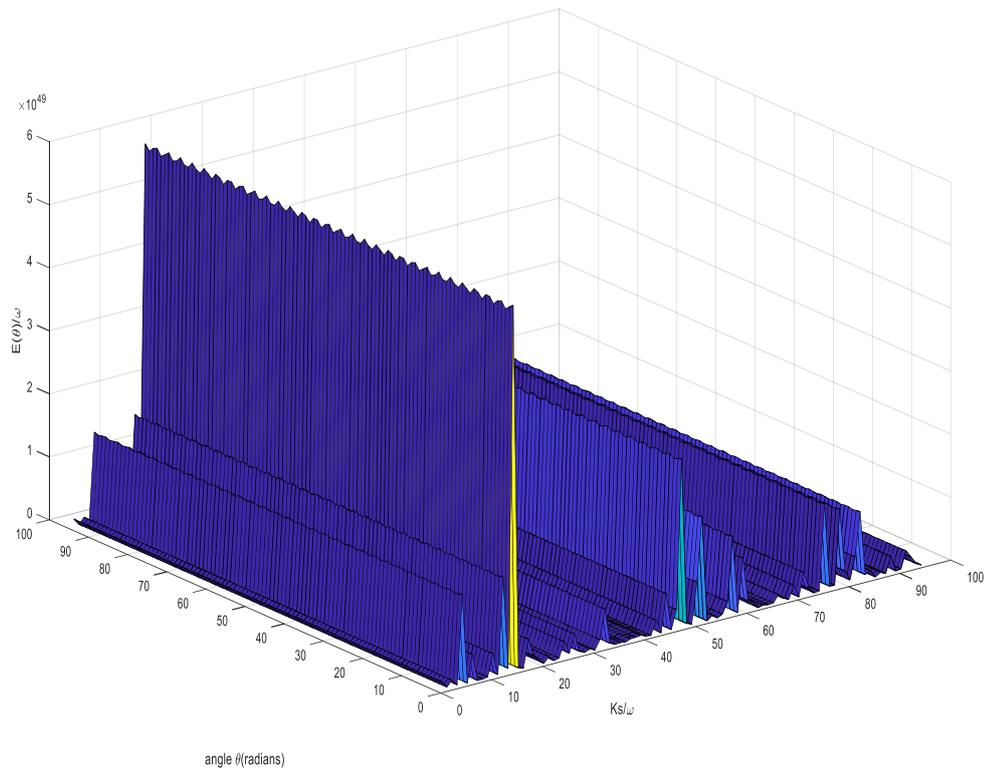

**Figure 3: 3-D plot of energy versus angle and stress induced anisotropy constant for**

$$\frac{D_1^{(4)}}{\omega} = 10, \qquad \frac{D_2^{(4)}}{\omega} = 5$$

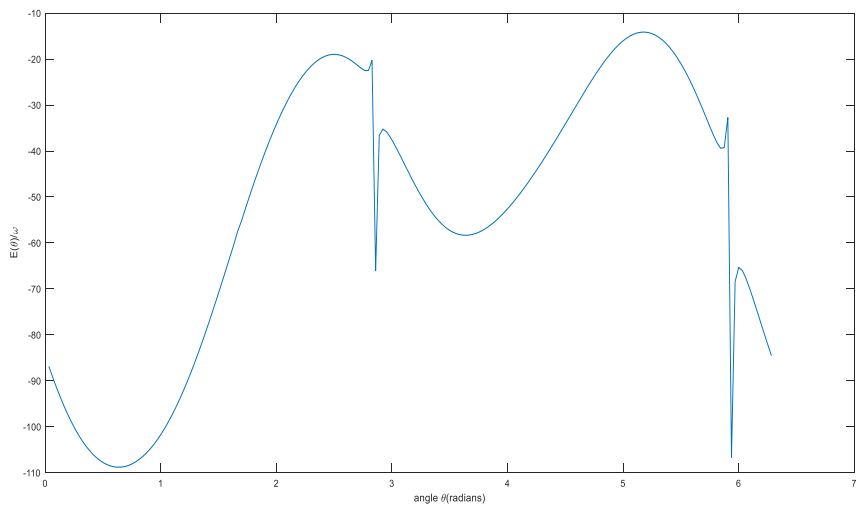

**Figure 4: Graph of energy versus angle for $\frac{K_s}{\omega} = 12$**

Figure 5 represents the 3D plot of total magnetic energy versus angle and stress induced magnetic anisotropy for $\frac{D_1^{(2)}}{\omega} = 100$, $\frac{D_2^{(2)}}{\omega} = 10$. Other energy parameters were kept at 10 for this simulation. The total magnetic energy does not considerably vary due to the increase of the second order magnetic anisotropy constant of the bottom spin layer.

.

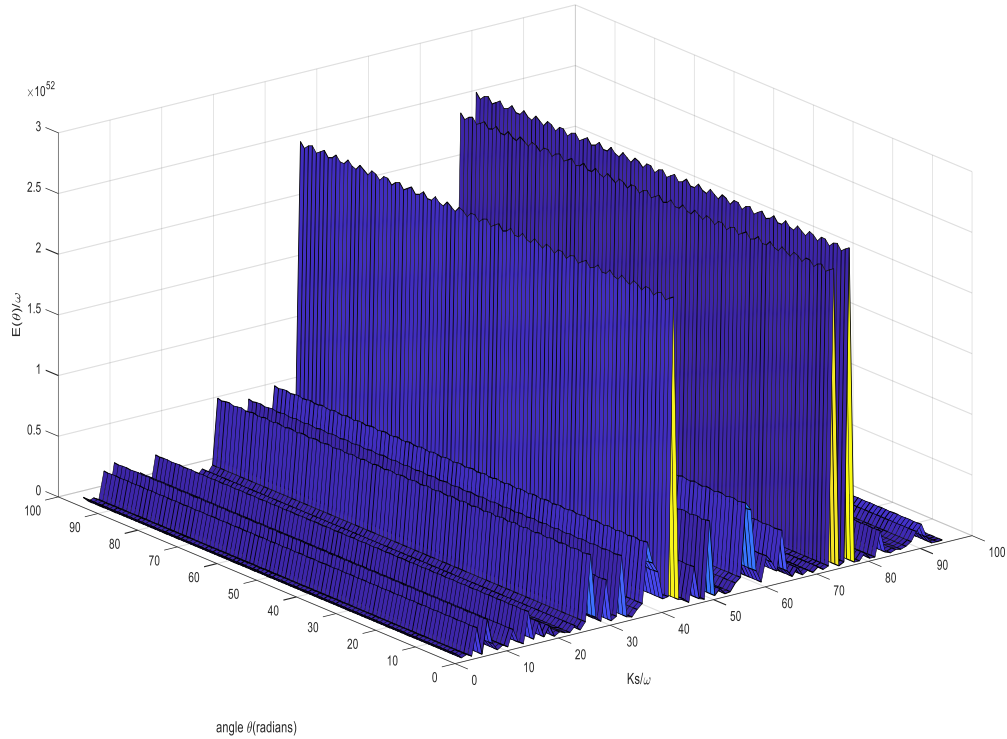

**Figure 5: 3-D plot of energy versus angle and stress induced anisotropy constant for**

$$\frac{D_1^{(2)}}{\omega} = 100, \qquad \frac{D_2^{(2)}}{\omega} = 10$$

## 4. Conclusion:

All the graphs between magnetic energy versus angle and stress induced anisotropy constant were plotted using fourth order perturbed Heisenberg Hamiltonian with all seven magnetic parameters for fcc structure with two spin layers. According to 3D plots the higher order of magnetic energy ($10^{51}$) was observed, when the fourth order magnetic anisotropy constant of top spin layer is higher than that of the bottom spin layer. Here the energy maximums were observed at $\frac{K_S}{\omega} = 12, 18, 28, 35$ and $90$. for $\frac{K_S}{\omega} = 12$, energy minimums were observed at 0.6283 and 3.581 radians and the energy maximum were found at 2.419 and 5.089 radians. The lower order of magnetic energy ($10^{49}$) was observed, when the fourth order magnetic anisotropy constant of top spin layer is less than that of the bottom spin layer. Here the energy maximums were observed at $\frac{K_s}{\omega} = 5, 15, 48, 52$ and 76. for $\frac{K_s}{\omega} = 15$, energy minimums were observed at 0.6283 and 3.644 radians and the energy maximum were found at 2.513 and 5.184 radians. In all cases the angle between consecutive magnetic easy and hard direction is not 90 degrees.

**References**

1. D. Zhao, Feng Liu, D.L. Huber and M.G. Lagally. Step-induced magnetic-hysteresis anisotropy in ferromagnetic thin films. *Journal of Applied Physics* 2002; 91(5): 3150.

2. M. Bentaleb, N. Aouad El and M. Saber. Phase Diagram of Spin 1/2 Ising Semi-Infinite and Film Ferromagnets with Spin 3/2 Overlayers. *Chinese Journal of Physics* 2002; 40(3): 307.

3. D. Spisak and J. Hafner. Theoretical study of FeCo/W (110) surface alloys. *Journal of Magnetism and Magnetic Materials.* 2005; 286: 386.

4. A. Ernst, M. Lueders, W.M. Temmerman, Z. Szotek and G. Van der Laan. Theoretical study of magnetic layers of nickel on copper; dead or alive? *Journal of Physics: Condensed matter.* 2000; 12(26): 5599.

5. P. Samarasekara and F.J. Cadieu. Magnetic and Structural Properties of RF Sputtered Polycrystalline Lithium Mixed Ferrimagnetic Films. *Chinese Journal of Physics* 2001; 39(6): 635-40.

6. P. Samarasekara and S.N.P. De Silva. Heisenberg Hamiltonian solution of thick ferromagnetic films with second order perturbation. *Chinese Journal of Physics* 2007; 45(2-I): 142.

7. P. Samarasekara and Udara Saparamadu. In plane oriented Strontium ferrite thin films described by spin reorientation. *Research & Reviews: Journal of Physics-STM Journals* 2013; 2(2): 12-16.

8. P. Samarasekara and E.M.P. Ekanayake. Ferromagnetic thin films with simple Cubic structure as described by fourth order perturbed Heisenberg Hamiltonian. *STM journals:Journal of Nanoscience, NanoEngineering & Applications* 2020; 10(3): 35-43.

9. P. Samarasekara and W.A. Mendoza. Effect of third order perturbation on Heisenberg Hamiltonian for non-oriented ultra-thin ferromagnetic films. *Electronic Journal of Theoretical Physics* 2010; 7(24): 197-210.

10. P. Samarasekara and B.I. Warnakulasooriya. Five layered fcc ferromagnetic films as described by modified second order perturbed Heisenberg Hamiltonian. *Journal of Science of the University of Kelaniya Sri Lanka* 2016; 11: 11-21.

11. N.U.S. Yapa, P. Samarasekara and Sunil Dehipawala. Variation of magnetic easy direction of bcc structured ferromagnetic films up to thirty spin layers. *Ceylon Journal of Science* 2017; 46(3): 93.

12. P. Samarasekara, M.K. Abeyratne and S. Dehipawalage. Heisenberg Hamiltonian with Second Order Perturbation for Spinel Ferrite Thin Films. *Electronic Journal of Theoretical Physics* 2009; 6(20): 345.

13. P. Samarasekara and S.N.P. De Silva. Heisenberg Hamiltonian solution of thick ferromagnetic films with second order perturbation. *Chinese Journal of Physics* 2007; 45(2-I): 142.

14. P. Samarasekara. Influence of third order perturbation on Heisenberg Hamiltonian of thick ferromagnetic films. *Electronic Journal of Theoretical Physics* 2008; 5(17): 227.

15. A. Hucht and K.D. Usadel. Reorientation transition of ultrathin ferromagnetic films. *Physical Review B* 1997; 55: 12309.

16. A. Hucht and K.D. Usadel. Theory of the spin reorientation transition of ultra-thin ferromagnetic films. *Journal of Magnetism and Magnetic Materials* 1999; 203(1): 88.

17. K.D. Usadel and A. Hucht. Anisotropy of ultrathin ferromagnetic films and the spin reorientation transition. *Physical Review B* 2002; 66: 024419.

18. M.S.M. Farhan and P. Samarasekara .Fourth Order Perturbed Heisenberg Hamiltonian with Seven Magnetic Parameters of bcc Structured Ferromagnetic Ultrathin Films. *Georgian Electronic Scientific Journals: Physics* 2022; 1(26): 29-45.

19. M.S.M. Farhan and P. Samarasekara. A solution of fourth order perturbed Heisenberg Hamiltonian with seven magnetic parameters. *STM journals: Journal of Nanoscience, NanoEngineering & Applications* 2022; 12(1): 1-13.